# Development of homogeneous and high-performance REBCO bulks with flexibility in shapes by the single-direction melt growth (SDMG) method

Takanori Motoki[1]*, Rempei Sasada[1], Takuma Tomihisa[1], Masaya Miwa[1],
Shin-ichi Nakamura[2] and Jun-ichi Shimoyama[1]

[1]*Department of Physical Sciences, Aoyama Gakuin University, 5-10-1 Fuchinobe, Chuo-ku, Sagamihara, Kanagawa 252-5258, Japan*
[2]*TEP Co. Ltd, 2-20-4 Kosuge, Katsushika-ku, Tokyo 124-0001, Japan*

E-mail: motoki@phys.aoyama.ac.jp

## Abstract

We have developed a single-direction melt growth method in which REBCO melt-textured bulks grow only vertically from a seed plate utilizing the difference in peritectic temperatures of REBCO. Entirely *c*-grown YBCO, DyBCO and GdBCO bulks with various sizes and shapes were successfully fabricated with high reproducibility. Disk-shaped bulks showed high trapped fields with almost concentric field distributions, reflecting homogeneous and boundaryless bulky crystal. In particular, a YBCO bulk with a 32 mm diameter trapped a high field more than 1 T at 77 K. Furthermore, rectangular and joined hexagonal REBCO bulks were successfully fabricated, showing designed field-trapping distributions reflecting their shapes through well-connected superconducting joints among bulks.

## 1. Introduction

REBa$_2$Cu$_3$O$_y$ (REBCO, RE: rare earth element) textured materials have been energetically developed owing to their high critical current density ($J_c$) under magnetic fields up to high temperatures, such as the temperature of liquid nitrogen, 77 K. Among them, REBCO melt-textured bulks have been widely studied especially for strong magnets using their high field-trapping ($B_T$) properties as high as ~17 T, due to the large persistent current circulating in bulks [1-3]. Both the uniformity and reproducibility of $B_T$ distributions as well as high $B_T$ of REBCO bulks are essential for extensive applications. Furthermore, bulks with various shapes such as rectangular, hexagonal, fan-shaped, and hollow cylindrical shapes, in addition to typical disk shapes are required for applications such as desktop NMR/MRI, bearings, undulator and motor systems [4-10]. In order to achieve biaxial texturing, melt growth using a small seed crystal placed on top of the precursor REBCO pellet has been generally adopted, which is well known as 'top-seeded melt-growth' (TSMG) or 'top-seeded infiltration growth' (TSIG) method. However, it is difficult to fabricate REBCO bulks with $B_T$ distributions clearly reflecting bulk shapes since top-seeded bulks are composed of *a*-growth and *c*-growth regions showing different crystallinity and critical current properties [11, 12]. In addition, fabrication of large bulks requires a very long time for both the crystal growth and oxygenation process, and random nucleation from the periphery part of the bulks is easily occurred as a result of the large supercooling degree. To address these issues, several attempts have been reported, such as the insertion of a buffer pellet, the addition of BaO$_2$ to decrease both the RE / Ba substitution and the peritectic temperature of REBCO, the modification of seeding, melt growth and oxygenation profiles, control of particle shapes and sizes of precursor powder, multi-seeding and intentional compositional gradient of elements along the radial direction of the bulk [13-21]. Regarding these problems, we have developed an innovative alternative to the top-seeded method, the 'single-direction melt growth (SDMG)' method [22]. In this method, large bulk plates cut from commercial TSMG REBCO bulks with relatively high peritectic temperature ($T_p$), such as EuBCO and GdBCO, are used as seed plates. REBCO bulks with $T_p$ lower than that of the seed plate are grown only along a vertical direction from the seed plate. Therefore, grown bulks consist of only a single grain region, and the crystal growth time does not depend on the radial size and shape of the bulks in principle. It is worth mentioning that seed plates can be reused many times after surface polishing, which is one of the advantages of high reproducibility. In the previous report, we succeeded in the fabrication of entirely $c$-grown YBCO bulks using SDMG method, which showed better concentrically cone-shaped $B_T$ distributions compared to those of TSMG processed bulks [22]. In the present study, we have attempted to fabricate entirely $c$-grown REBCO bulks with various RE elements, sizes and shapes by the SDMG method.

## 2. Experimental

An overview of the SDMG process is summarized in Fig. 1. Randomly oriented (RE')BCO sintered bulks placed on a large seed plate cut from commercial TSMG (RE")BCO bulks with (001) surface are epitaxially melt grown from the interface to the top, where (RE")BCO must have a $T_p$ higher than that of (RE')BCO. In other words, Y or heavy rare earth elements and light rare earth elements are suitable for RE' and RE", respectively. Hereafter, when it is necessary to avoid confusion, the notations RE' and RE" will be used to distinguish RE of grown bulks and seed plates, respectively.

We selected RE' = Y, Dy, Gd and RE" = Gd, Eu, respectively. Precursor powders with a mixture of RE'BCO and RE'$_2$BaCuO$_5$ (RE'211) with a molar ratio of 7 : 3 for RE' = Y, Dy and 7.5 : 2.5 for RE' = Gd, respectively, were produced by *TEP Co.* 10 wt% Ag$_2$O was mixed with the precursor to decrease $T_p$ and to enhance the mechanical strength of the grown bulks. 0.5 wt% CeO$_2$ was also added to suppress the grain growth of RE'211. The precursor powder was pressed into disks with various diameters or other shapes under a uniaxial pressure of ~100 MPa. Precursor RE'BCO bulks were pre-melted in air for ~2 h for densification, followed by surface-polishing. The densified RE'BCO bulks were placed on a RE"BCO seed plate with (001) surface sliced from commercial TSMG bulks (Gd-QMG® or Eu-QMG®) made by *Nippon Steel Co*. A slurry consisting of ethanol and precursor powder was coated as a buffer layer at the interface between the bulks and the seed plate to ensure uniform contact. SDMG was carried out in air inside a standard box furnace without intentional temperature gradient. It is important to achieve the partially melting state of the RE'BCO bulk and to avoid deformations of the RE"BCO seed plate that the maximum holding temperature must be above the $T_p$ of the RE'BCO bulk and below the $T_p$ of the RE"BCO seed plate. The $T_p$s of Ag-added YBCO, DyBCO, GdBCO and EuBCO are ~980, ~990, ~1010 and ~1030°C, respectively, considering that the addition of Ag lowers $T_p$s by approximately 20°C [23]. The optimized temperature profiles for RE'BCO bulks with different RE elements and the setting geometry are shown in Fig. 2 and table 1. After SDMG, grown bulks were separated from the seed plate by cutting using a diamond saw, followed by reductive annealing at 800–850°C under flowing 1%O$_2$ (Ar balanced) gas for 24 h to suppress the level of RE/Ba substitution. Finally, oxygenation in a tube furnace was carried out down to 450, 425 and 350°C for YBCO, DyBCO and GdBCO, respectively, for more than 200 h under flowing pure oxygen to achieve the carrier optimally doped state. Thus, SDMG processed entirely $c$-grown RE'BCO bulks with various diameters and shapes can be directly obtained on a [001] oriented RE"BCO seed plate. The glossy faces typical of melt-textured bulks can be clearly observed to be near the top surface for all of the prepared bulks, suggesting sufficient crystal growth of the entire bulks. It is noteworthy that there is almost no failure in crystal growth using the SDMG method under the same temperature profiles indicated in Fig. 2 and Table 1 as long as the placed bulks are in uniform contact with the seed plate, representing high reproducibility of the SDMG method. For comparison, two DyBCO bulks were prepared under the identical condition by the conventional TSMG method using a NdBCO single crystal as a seed crystal. Detailed conditions of TSMG are described in our previous paper [24]. In addition to standard disk-shaped bulks, we have demonstrated direct preparation of rectangular and hexagonal-shaped bulks using the SDMG method, where metal dies with a side length of 16 and 9.2 mm were used for pelletization of the rectangular and the hexagonal bulks, respectively.

For a SDMG YBCO bulk grown on a GdBCO seed plate, crystallinity of the seed plate and the grown bulk was evaluated by the x-ray diffraction (XRD, PANalytical X'PERT PRO MRD) and the bulk/seed interface was observed using a polarized optical microscopy, secondary electron microscopy (SEM, ZEISS ULTRA55), and 200 keV transmission electron microscopy (TEM, JEOL JEM-2100) along with elemental analysis using energy dispersive X-ray spectroscopy (EDX) attached to the SEM and TEM. Small samples for the TEM obsevation were processed and picked up by the focused ion beam apparatus (FIB, Hitachi High-Tech MI4050).

$B_T$ distributions ~0.5 mm above the polished surface of the bulk (seed plate side) were examined by scanning a Hall

probe at 77 K. Magnetization was carried out by field cooling up to 2 T at 77 K. The $B_{T, max}$ is the highest $B_T$ value was measured just above the surface of the bulks.

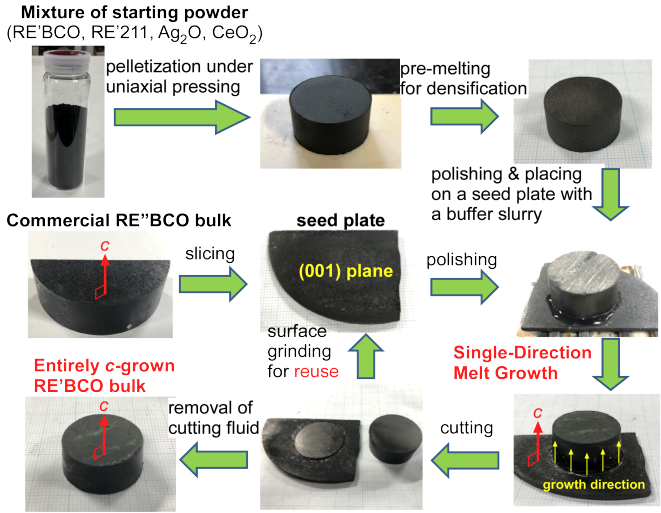

**Figure 1**

Experimental overview of the single-direction melt growth (SDMG) method.

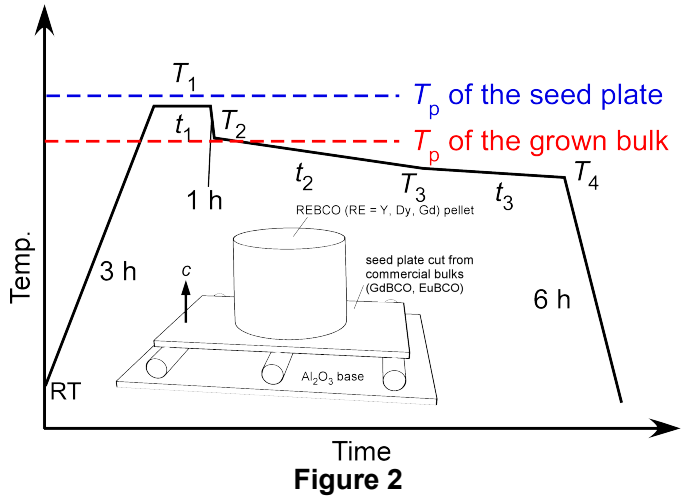

**Figure 2**

Temperature profiles and schematic configuration of the SDMG method.

**Table 1**

REBCO combinations for bulks and seed plates prepared in this study along with the temperature profiles of the SDMG processes corresponding to Fig.2. Note that all SDMG profiles are decided for Ag-added REBCO bulks and seed plates.

| Grown pellet | Seed plate | Temperature / °C | | | | Time / hour | | |
|---|---|---|---|---|---|---|---|---|
| | | $T_1$ | $T_2$ | $T_3$ | $T_4$ | $t_1$ | $t_2$ | $t_3$ |
| YBCO | GdBCO | 990 | 970 | 950 | 950 | 2 | 80 | 0 |
| DyBCO | EuBCO | 1015 | 990 | 958 | 940 | 2 | 80 | 60 |
| GdBCO | EuBCO | 1015 | 1015 | 983 | 971 | 0 | 80 | 40 |

## 3. Results and Discussion

First, crystallinities of a GdBCO seed plate and a grown YBCO bulk were evaluated by XRD. Figs. 3 show the pole figures of GdBCO (103) for the seed plate (a) and YBCO (103) for the polished surface on the seed side of the grown bulk (b), respectively. In both figures, four symmetrical sharp peaks were observed and full width at half maximum (FWHM) values of $\Delta\phi$ averaged over four peaks are ~3.5° and ~3.2° for the GdBCO seed and the grown YBCO, respectively. The crystallinity of the grown YBCO was found to be almost unchanged to that of the GdBCO seed plate.

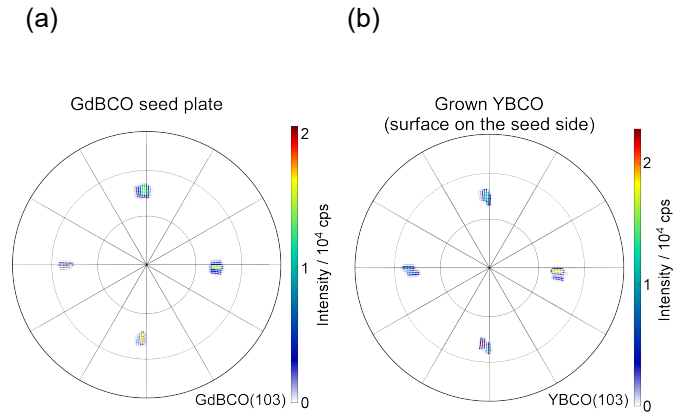

**Figure 3 (a),(b)**

(103) pole figures of the GdBCO seed plate (a) and the surface on the seed side of the grown YBCO bulk using the SDMG method (b).

Second, the microstructure of the interface at the SDMG grown YBCO/GdBCO seed plate was observed to evaluate the interfacial cleanliness and the degree of interdiffusion of RE elements across the interface. Fig. 4 (a) shows a polarized optical microscope image of a wide range of the cross-sectional bulk/seed interface. The interface is clean and undistinguishable without coarse impurities and/or cracks, while there are clear differences in the size and distribution of Ag and voids between the seed plate and bulk region. SEM and TEM observations were carried out for the region corresponding to the dashed line in Fig. 4 (a) and the micro area that includes the interface, respectively. Figs. 4 (b) and (c) show SEM and TEM images around the YBCO/GdBCO interface along with elemental mappings of Gd and Y. As for TEM observation, selected area electron diffraction (SAED) patterns were also evaluated at three positions across the interface (corresponding to positions A, B, and C). In both elemental mappings with different magnifications in Figs. 4. (b) and (c), the clear change in the element distributions can be seen at the interface, indicating the very low level of interdiffusion of RE elements after SDMG. Therefore, the seed plate can be reused many times after polishing at least a few micrometers. In fact, we have confirmed that SMDG bulks can be reproducibly obtained even after reusing a ~3 mm thick seed plate several times. As for the TSMG or TSIG method, the seed crystals are usually single-use, whereas in this SDMG method, it is one of the significant advantages that the seed plates can be used repeatedly. The SAED patterns in Fig. 4 (c) show that the grown bulk is well biaxially oriented, inheriting the orientation of the seed plate, which is consistent with the four sharp peaks in the pole figures as shown in Figs. 3.

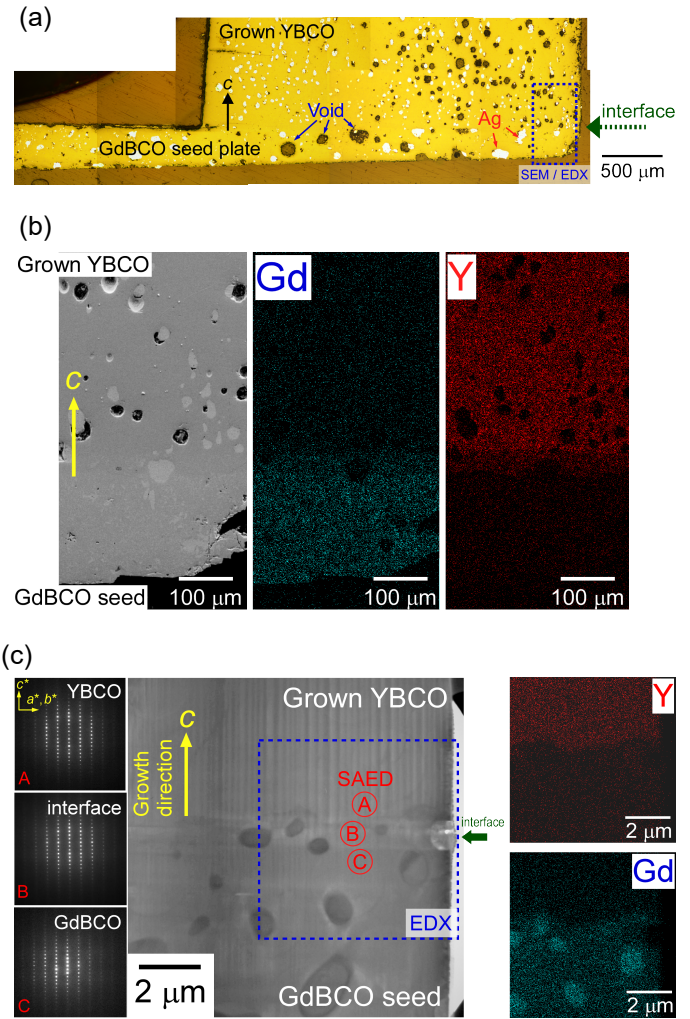

**Figure 4 (a), (b), (c)**

Optical (a), SEM (b) and TEM (c) images around the interface between the SDMG YBCO bulk and the GdBCO seed plate. Elemental mappings of Y and Gd corresponding to the area surrounded by the dashed lines are indicated in (b) and (c). Selected area electron diffraction patterns of three different positions across the interface are also shown in (c).

Fig. 5 shows two-dimensional contour mappings of $B_T$ distributions at 77 K of SDMG processed GdBCO (#S1), DyBCO (#S2 and #S3), and YBCO (#S4) bulks arranged in order of increasing diameter, and TSMG processed DyBCO (#T1 and #T2) bulks prepared under the identical condition. Thereafter, 'mm$\phi$' will be used to denote the bulk diameter. Although the TSMG-processed bulks showed rectangular-like $B_T$ distributions with poor reproducibility despite being fabricated under identical condition, all SDMG bulks showed uniform $B_T$ distributions with high circularity regardless of RE elements. They also exhibited high $B_{T,max}$ which increases as an increase in bulk diameters. High $B_{T,max}$ values of 0.79 T and 1.04 T at 77 K were recorded by SDMG processed DyBCO and YBCO bulks with 20.8 and 32.0 mm$\phi$, respectively. These values were comparable with the reported 24.15 mm$\phi$ GdBCO bulk exhibiting the highest $B_T$ properties ever, *i.e.* $B_T$ >17 T at ~30 K, showed $B_{T,max}$ of ~0.9 T at 77 K [3]. The quantitative evaluation of the degree of roundness will be discussed later in this paper. Since the SDMG-processed bulks have no *a/c*-growth grain boundaries with homogeneous radial $J_c$ distributions unlike typical top-seeded bulks, uniform and concentric $B_T$ distributions are achieved through field-cooling magnetization. Moreover, the $B_T$ distributions can also be improved for bulks magnetized by the pulsed field method because selective magnetic field invasion through the grain sector region (GSR) prior to grain boundary region (GSB) reported in TSMG bulks [25, 26] does not occur in case of the SMDG bulks.

Finally, direct fabrication of bulks with various shapes was attempted using the SDMG method. Fig. 6 shows the appearances and $B_T$ distributions of a rectangular YBCO bulk (#S5) and a joined hexagonal DyBCO bulk (#S6), respectively. The #S6 DyBCO bulk was grown from three hexagonal pellets just placed adjacent to each other on a EuBCO seed plate. It was demonstrated that the $B_T$ distributions clearly reflect the shapes of the bulks, indicating the high uniformity in the *ab* plane of these complex-shaped bulks. Furthermore, little or no decrease in $B_T$ at the boundary of the original pellets was observed for the #S6 joined hexagonal bulk, suggesting that well-connected superconducting joints among the three pellets can be easily realized by the SDMG method through simply placing the pre-melted pellets in contact with each other. Further investigation on the formation of superconducting joints through SDMG is currently underway. In addition, we have succeeded in the direct growth of hollow-cylindrical bulks, for which a few studies have been reported using the TSMG or TSIG method [27, 28], and these results will be reported in a future paper.

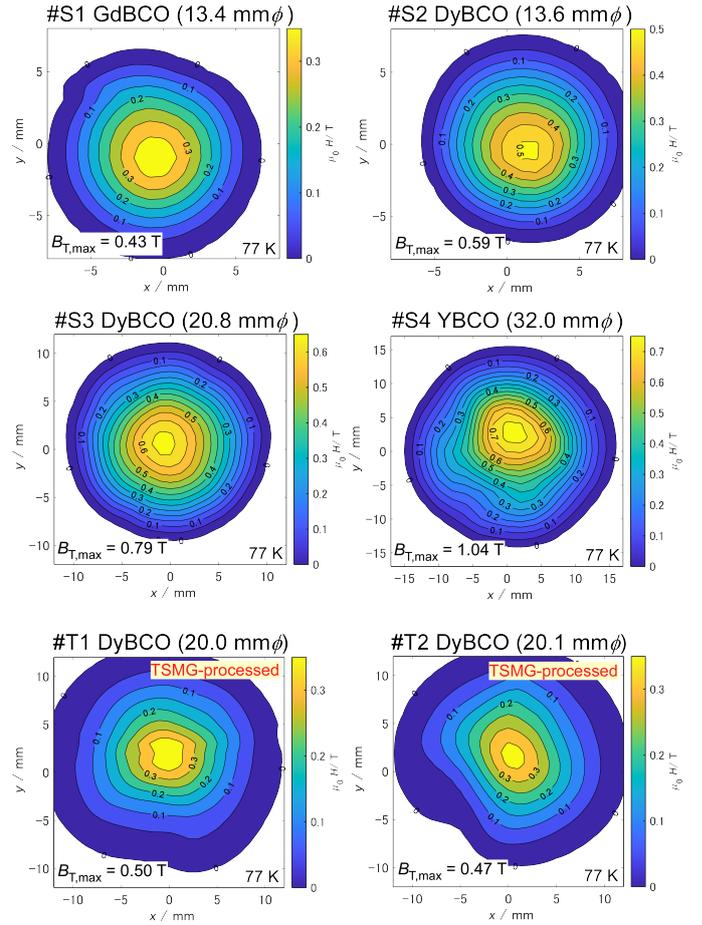

**Figure 5**
Two-dimensional $B_T$ distributions at 77 K of SDMG processed GdBCO (#S1), DyBCO (#S2 and #S3), YBCO (#S4) and TSMG processed DyBCO (#T1 and #T2) bulks measured ~0.5 mm above the surface of the bulks. $B_{T,max}$ exhibits the highest $B_T$ value when the Hall probe is in contact with the surface of the bulks.

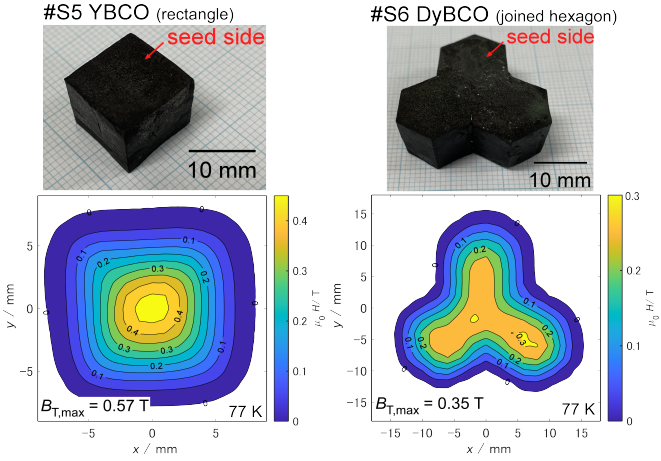

**Figure 6**

Photographs and $B_T$ distributions at 77 K of rectangular-shaped YBCO (#S5) and joined hexagonal-shaped DyBCO (#S6) bulks directly prepared using the SDMG method.

In order to evaluate the degree of roundness of $B_T$ distributions for the prepared SDMG and TSMG bulks quantitively, we adopted the circularity, $c$, and ellipticity, $e$, as a scale for comparison of homogeneity. Specifically, when the $B_T$ distribution gets closer to a perfect circle, $c$ approaches to one while $e$ approaches to zero. For detailed definitions, refer to our previous paper [22]. The averaged values of $c$ and $e$ were calculated from $B_T$ contours with intervals of 0.02 T. The calculated results for $c$ and $e$ are summarized in Table 2 along with the preparation methods, shapes, bulk sizes, and $B_{T,max}$. All SDMG-processed bulks in this study exhibited superior concentric indexes, i.e. higher circularity and lower ellipticity, regardless of the RE elements of the bulks. It should be noted that the reproducibility and roundness of the $B_T$ distributions are slightly low for the large bulks as seen in #S4 YBCO 32.0 mm$\phi$ bulk. This may be due to the fact that a uniform contact between the seed and the pellet was not sufficiently achieved. More careful surface polishing of the seed plates and optimization of coating conditions of buffer layer will solve this problem.

## 4. Conclusions

In conclusion, we have developed a 'single-direction melt growth (SDMG)' method for the fabrication of entirely $c$-grown REBCO melt-textured bulks. In this method, REBCO bulks can be grown directly regardless of bulk diameters and shapes, since bulks are grown only in the vertical direction from a large seed plate sliced from commercial REBCO bulks having slightly higher peritectic temperatures, such as EuBCO and GdBCO. Microstructural observation revealed that the bulk/seed interface is clean and that there is very low level of interdiffusion of RE elements, allowing multiple reuses of seed plates with only a little surface polishing. REBCO bulks with various sizes and shapes were successfully grown using the SDMG method for RE = Y, Dy and Gd. Both high circularity and high field trapping properties, i.e. the trapped field value of ~1 T at 77 K, were achieved for the SDMG bulks. In addition, we have succeeded in the direct growth of bulks with various shapes which exhibited trapped field distributions clearly reflecting their complex shapes including the joined hexagon through achieving well-connected superconducting joints among adjacent multiple bulks. The SDMG method is very promising for fabrication of high-performance melt-textured bulks with high scalability and flexibility of shapes.

**Table 2**

Summary of the grown REBCO, preparation method, bulk shape, size, $B_{T,max}$, circularity, and ellipticity for the prepared bulks in this study. $\phi$ and $t$ represent the diameter and thickness of the bulks, respectively.

| No. | REBCO | Method | Shape | Size / mm | $B_{T,max}$ (77 K) / T | Circularity, $c$ | Ellipticity, $e$ |
|---|---|---|---|---|---|---|---|
| #S1 | GdBCO | SDMG | Disk | 13.4$\phi$ × 8.2$^t$ | 0.43 | 0.983 | 0.050 |
| #S2 | DyBCO | SDMG | Disk | 13.6$\phi$ × 7.7$^t$ | 0.59 | 0.988 | 0.042 |
| #S3 | DyBCO | SDMG | Disk | 20.8$\phi$ × 9.9$^t$ | 0.79 | 0.991 | 0.042 |
| #S4 | YBCO | SDMG | Disk | 32.0 $\phi$ × 12.6$^t$ | 1.04 | 0.965 | 0.100 |
| #T1 | DyBCO | TSMG | Disk | 20.0 $\phi$ × 13.1$^t$ | 0.50 | 0.933 | 0.140 |
| #T2 | DyBCO | TSMG | Disk | 20.1 $\phi$ × 13.2$^t$ | 0.47 | 0.957 | 0.197 |
| #S5 | YBCO | SDMG | Rectangle | 13.1×13.1×11.3$^t$ | 0.57 | - | - |
| #S6 | DyBCO | SDMG | Joined hexagon | 7.7$^{(each\ side)}$ × 10.8$^t$ | 0.35 | - | - |

## Acknowledgements

This work was partially supported by JSPS KAKENHI, grant number 19K05006, Japan. TEM observation and elemental analysis in this work were performed at the Center for Instrumental Analysis, College of Science and Engineering, Aoyama Gakuin University.

## References

[1] Muralidhar M and Murakami M 2000 Superconducting properties of (Nd, Eu, Gd)-123 *Physica C* **341-348** 2431–2
[2] Tomita M and Murakami M 2003 High-temperature superconductor bulk magnets that can trap magnetic fields of over 17 tesla at 29 K *Nature* **421** 517–20
[3] Durrell J H, Dennis A R, Jaroszynski J, Ainslie M D, Palmer K G B, Shi Y H, Campbell A M, Hull J, Strasik M, Hellstrom E E and Cardwell D A 2014 A trapped field of 17.6 T in melt-processed, bulk Gd-Ba-Cu-O reinforced with shrink-fit steel *Supercond. Sci. Technol.* **27** 082001
[4] Zhou D, Izumi M, Miki M, Felder B, Ida T and Kitano M 2012 An overview of rotating machine systems with high-temperature bulk superconductors *Supercond. Sci. Technol.* **25** 103001
[5] Durrell J H, Ainslie M D, Zhou D, Vanderbemden P, Bradshaw T, Speller S, Filipenko M and Cardwell D A 2018 Bulk superconductors: A roadmap to applications *Supercond. Sci. Technol.* **31** 103501
[6] Nakamura T, Tamada D, Yanagi Y, Itoh Y, Nemoto T, Utumi H and Kose K 2015 Development of a superconducting bulk magnet for NMR and MRI *J. Magn. Reson.* **259** 68–75
[7] Werfel F N, Floegel-Delor U, Rothfeld R, Riedel T, Goebel B, Wippich D and Schirrmeister P 2012 Superconductor bearings, flywheels and transportation *Supercond. Sci. Technol.* **25** 014007
[8] Hlásek T and Plecháček V 2015 Trapped field in different shapes of RE-Ba-Cu-O single grains for the use in production of superconducting bearings *IEEE Trans. Appl. Supercond.* **25** 25–8
[9] Kii T, Kinjo R, Kimura N, Shibata M, Bakr M A, Choi Y W, Omer M, Yoshida K, Ishida K, Komai T, Shimahashi K, Sonobe T, Zen H, Masuda K and Ohgaki H 2012 Low-temperature operation of a bulk HTSC staggered array undulator *IEEE Trans. Appl. Supercond.* **22** 11–4
[10] Terao Y, Akada W and Ohsaki H 2019 Design and Comparison of Interior Permanent Magnet Synchronous Motors Using Different Bulk Superconductor Arrangements *IEEE Trans. Appl. Supercond.* **29** 1–5
[11] Nakashima T, Shimoyama J, Honzumi M, Tazaki Y, Horii S and Kishio K 2005 Relationship between critical current properties and microstructure in cylindrical RE123 melt-solidified bulks *Physica C* **426–431** 720–5
[12] Diko P 2006 Microstructural limits of TSMG REBCO bulk superconductors *Physica C* **445–448** 323–9
[13] Cardwell D A, Shi Y and Namburi D K 2020 Reliable single grain growth of (RE)BCO bulk superconductors with enhanced superconducting properties *Supercond. Sci. Technol.* **33** 024004
[14] Namburi D K, Shi Y and Cardwell D A 2021 The processing and properties of bulk (RE)BCO high temperature superconductors: Current status and future perspectives *Supercond. Sci. Technol.* **34** 053002
[15] Babu N H, Shi Y, Iida K and Cardwell D A 2005 A practical route for the fabrication of large single-crystal (RE)–Ba–Cu–O superconductors *Nat. Mater.* **4** 476–80
[16] Motoki T, Yanai Y, Nunokawa K, Gondo S, Nakamura S and Shimoyama J 2020 Breakthrough in the reduction of oxygen-Annealing time for REBCO melt-Textured bulks under an oxygen atmosphere containing water vapor *Supercond. Sci. Technol.* **33** 034008
[17] Namburi D K, Shi Y, Dennis A R, Durrell J H and Cardwell D A 2018 A robust seeding technique for the growth of single grain (RE)BCO and (RE)BCO-Ag bulk superconductors *Supercond. Sci. Technol.* **31** 044003
[18] Yang P, Yang W, Zhang L and Chen L 2018 Novel configurations for the fabrication of high quality REBCO bulk superconductors by a modified RE + 011 top-seeded infiltration and growth process *Supercond. Sci. Technol.* **31** 085005
[19] Pavan S, Naik K, Muralidhar M, Koblischka M R, Koblischka-veneva A, Oka T and Murakami M 2019 Novel method of tuning the size of $Y_2BaCuO_5$ particles and their influence on the physical properties of bulk $YBa_2Cu_3O_{7-\delta}$ superconductor *Appl. Phys. Express* **12** 063002
[20] Sawamura M, Morita M and Hirano H 2003 Enlargement of bulk superconductors by the MUSLE technique *Physica C* **392–396** 441–5
[21] Nariki S, Teshima H and Morita M 2016 Performance and applications of quench melt-growth bulk magnets *Supercond. Sci. Technol.* **29** 034002
[22] Motoki T, Yanai Y, Nunokawa K and Shimoyama J 2020 Fabrication of high-performance $YBa_2Cu_3O_y$ melt-textured bulks with selective grain growth *Appl. Phys. Express* **13** 093002
[23] Murakami M, Sakai N, Higuchi T and Yoo S I 1996 Melt-processed light rare earth element-Ba-Cu-O *Supercond. Sci. Technol.* **9** 1015–32
[24] [previous TSMG] Setoyama Y, Shimoyama J, Yamaki S, Yamamoto A, Ogino H, Kishio K and Awaji S 2015 Systematic change of flux pinning in (Dy,RE)123 and (Y,RE)123 melt-solidified bulks with unit cell orthorhombicity *Supercond. Sci. Technol.* **28** 015014
[25] Ainslie M D, Fujishiro H, Ujiie T, Zou J, Dennis A R,

Shi Y H and Cardwell D A 2014 Modelling and comparison of trapped fields in (RE)BCO bulk superconductors for activation using pulsed field magnetization *Supercond. Sci. Technol.* **27** 065008

[26] Oka T, Hara K, Ogawa J, Fukui S, Sato T, Yokoyama K, Murakami A and Langer M 2016 Selective Magnetic Field Invasion into HTS Bulk Magnets in Pulse Field Magnetizing Processes *IEEE Trans. Appl. Supercond.* **26** 3–6

[27] Werfel F N, Floegel-Delor U, Riedel T, Goebel B, Rothfeld R, Schirrmeister P and Wippich D 2013 Large-scale HTS bulks for magnetic application *Physica C* **484** 6–11

[28] Yang P T, Yang W M and Chen J L 2017 Fabrication and properties of single domain GdBCO superconducting rings by a buffer aided Gd+011 TSIG method *Supercond. Sci. Technol.* **30** 085003